\documentstyle[12pt]{article}
\renewcommand{\theequation}
{\arabic{section}.\arabic{equation}}
\newcommand{\eqreset}{\setcounter{equation}{0}}
\newtheorem{theorem}{Theorem}[section]

\newtheorem{lemma}[theorem]{Lemma}
\begin{document}
\title{{\bf Solutions of  Discretized Affine Toda Field Equations 
for $A_{n}^{(1)}$, $B_{n}^{(1)}$, $C_{n}^{(1)}$,\\ $D_{n}^{(1)}$, 
 $A_{n}^{(2)}$  and $D_{n+1}^{(2)}$ }}
\author{Zengo Tsuboi \\ Institute of Physics, University of Tokyo \\
 Komaba 3-8-1, Meguro-ku, Tokyo 153}
\maketitle
\begin{abstract}
It is known that a family of
 transfer matrix functional equations, the T-system, 
can be compactly written in terms of the Cartan matrix of
 a simple Lie algebra.
We formally replace this Cartan matrix of a simple Lie 
algebra with that of an affine Lie algebra, and then we 
obtain a system of functional equations different from 
the T-system.
 It may be viewed as an $X_{n}^{(a)}$ type affine 
Toda field equation on discrete space time. 
We present, for $A_{n}^{(1)}$, $B_{n}^{(1)}$, $C_{n}^{(1)}$, 
$D_{n}^{(1)}$, $A_{n}^{(2)}$ and $D_{n+1}^{(2)}$ ,
its solutions in terms of determinants or Pfaffians.
\end{abstract}
 \hspace{45pt}\\
 solv-int / 9610011 \\ 
Journal-ref: J. Phys. Soc. Jpn. {\bf 66} (1997) 3391-3398 
\newpage
\eqreset
\section{Introduction}
Kuniba {\it et al.} proposed \cite{KNS1},
 a class of functional equations, the T-system.
 It is a family of functional equations satisfied by a set of commuting transfer
  matrices $\{T_{m}^{(a)}(u)\}$ of solvable lattice models associated with any 
quantum affine algebras $U_{q}(X_{r}^{(1)})$. Here, the transfer
 matrix $T_{m}^{(a)}(u)$ is defined as a trace of monodromy matrix 
 on auxiliary space labeled by $a$ and $m$ $(m \in {\bf Z}, 
u \in {\bf C}, a\in \{1, 2, \dots, r \})$. 
 
It was pointed out \cite{KOS} that the T-system is not only a family of 
transfer matrix 
functional equations but also an example of Toda field equation 
\cite{LS,MOP} on discrete space time. Solving the T-system
  recursively, we obtained 
 solutions $\{ T_{m}^{(a)}(u)\}$ \cite{KNH,KOS,TK} 
 with the following properties:
\begin{enumerate}
\item They are polynomials of the initial values $T_{1}^{(1)}(u+{\rm shift}), 
\dots , T_{1}^{(r)}(u+{\rm shift})$.   
\item They have determinant or pfaffian expressions.
\end{enumerate}
  
Using the Cartan matrix of a simple Lie algebra $X_{r}$, 
the T-system is written as a compact form \cite{KNS2}. 
In this sense, the T-system is
 considered to be a class of functional equations associated with Dynkin 
 diagrams. Then the following natural question arise. \\
{\it If we formally replace this Cartan matrix of a simple Lie 
algebra with that of a Kac-Moody Lie algebra,  then we obtain a system of
 functional equations different from the T-system. 
Can one construct solutions of these equations, 
which have similar properties as 
those of the T-system ?}
\\  
The purpose of this paper is to answer this question. So far,  
we have succeeded to construct solutions for 
 affine Lie algebras $A_{n}^{(1)}$, $B_{n}^{(1)}$,
 $C_{n}^{(1)}$, $D_{n}^{(1)}$, $A_{n}^{(2)}$ and 
$D_{n+1}^{(2)}$, modifying the expressions 
of solutions \cite{KOS,KNH,TK} of the  
 T-system.

The functional equations which we deal with in this paper 
 can be compactly written as follows: 
\begin{eqnarray}
T_{m}^{(a)}(u+\frac{1}{t_a}) T_{m}^{(a)}(u-\frac{1}{t_a})
=T_{m+1}^{(a)}(u) T_{m-1}^{(a)}(u)
+\prod_{b=0}^{r} {\cal T}(a,b,m,u)^{I_{ab}} \label{master}
\end{eqnarray}
Here, 
\begin{eqnarray}
{\cal T}(a,b,m,u)&=&T_{t_b m/t_a}(u) \quad {\rm for } \quad 
\frac{t_b}{t_a}=1,2,3, \\ 
&=&T_{\frac{m}{2}}^{(b)}(u+\frac{1}{2}) T_{\frac{m}{2}}^{(b)}(u-\frac{1}{2})
T_{\frac{m+1}{2}}^{(b)}(u) T_{\frac{m-1}{2}}^{(b)}(u) 
\ {\rm for } \ 
\frac{t_b}{t_a}=\frac{1}{2}, \\
&=&T_{\frac{m}{3}}^{(b)}(u+\frac{2}{3}) T_{\frac{m}{3}}^{(b)}(u)
T_{\frac{m}{3}}^{(b)}(u-\frac{2}{3})
T_{\frac{m+2}{3}}^{(b)}(u)  \\ 
&\times &T_{\frac{m-1}{3}}^{(b)}(u-\frac{1}{3})
T_{\frac{m-1}{3}}^{(b)}(u+\frac{1}{3})
T_{\frac{m-2}{3}}^{(b)}(u) T_{\frac{m+1}{3}}^{(b)}(u-\frac{1}{3})
T_{\frac{m+1}{3}}^{(b)}(u+\frac{1}{3}) 
\nonumber \\ 
&& \quad {\rm for } \quad 
\frac{t_b}{t_a}=\frac{1}{3} .\nonumber 
\end{eqnarray}
where $a \in \{0,1,2,\dots,r \}$; $T_{0}^{(a)}(u)=1$; $T_{m}^{(a)}(u)=1$
 ( if $m \notin Z$ ); 
$I_{ab}=2\delta_{ab}-B_{ab} (I_{ab}=1$,
 if $a$ th node of Dynkin diagram is connected with $b$ th node of it;
  $I_{ab}=0$,  
if $a$ th node of Dynkin diagram is disconnected with $b$ th node of it); 
$B_{ab}=C_{ab}\frac{t_b}{t_{ab}}$; 
$C_{ab}=\frac{2(\alpha_a|\alpha_b)}{(\alpha_a|\alpha_a)}$ ($C_{ab}$: Cartan 
matrix of affine Lie algebra; $\alpha_a$: simple root); 
$t_a= \frac{2}{(\alpha_a|\alpha_a)}; t_{ab}=max(t_a,t_b).$
This functional equation resembles the T-system except the 
0 th component $T_{m}^{(0)}(u)$.
However, unlike the T-system,
 the solutions of this functional equation (\ref{master})
 have nothing to do with 
the solvable lattice models to the author\symbol{39}  knowledge. 
Moreover, as is mentioned in section2, our functional equation 
 may be viewed as an example of discretized affine Toda field equation. 
 In this sense, the functional equation (\ref{master}) should be viewed as 
  the generalization of 
 a discretized Toda equation rather than that of the T-system. 
It has beautiful structure as an example of discretized soliton equation 
\cite{AL1,AL2,DJM1,DJM2,DJM3,DJM4,DJM5,H1,H2,H3,H4,HTI,S1,S2,Wa}.
Discrete equations have richer contents than their continuum counterparts.  
In fact, our new solutions here are not so-called 
soliton solutions which have continuum 
counterparts but the ones whose matrix sizes depend on the parameter 
$m$ and thus peculiar to difference equations.
 Solving the functional equation (\ref{master}) recursively, 
  we can express $T_{m}^{(a)}(u)$ as a 
  polynomial of $T_{1}^{(0)}(u+{\rm shift}), \dots 
  T_{1}^{(r)}(u+{\rm shift}) $. Combining the similar expressions of the 
  solutions \cite{KNH,KOS,TK} to the T-system,
   we present, for $A_{n}^{(1)}$, 
$B_{n}^{(1)}$, $C_{n}^{(1)}$, $D_{n}^{(1)}$, $A_{n}^{(2)}$ and 
$D_{n+1}^{(2)}$, solutions of eq.(\ref{master}) 
in terms of determinants or pfaffians. 

For simple Lie algebras $B_{r}$ and $D_{r}$, there are at least two 
types of expressions of the solutions to the T-system, 
which are equivalent.
The one is expressed {\it explicitly} by polynomial of the fundamental 
polynomials $T_1^{(1)},\dots,T_1^{(r)}$ and is given as a 
determinant or a pfaffian 
whose non-zero matrix elements distribute sparsely \cite{KNH}. 
The other is expressed by polynomial of 
  auxiliary transfer matrix ${\cal T}^{a}(u)$
 (or `dress function' in the analytic Bethe ansatz), and thus 
 {\it implicitly} by polynomial of the fundamental polynomials 
$T_1^{(1)},\dots,T_1^{(r)}$ and is given as a 
determinant or a pfaffian 
whose non-zero matrix elements distribute densely \cite{KOS,TK}. 
On constructing the solution of the functional equation (\ref{master}), 
we use only the dense type expression of the solution to the  
T-system. The boundary condition to the function 
 ${\cal T}^{a}(u)$ for eq.(\ref{master}) 
 is quite different from that for the T-system.
 As a result, in spite of the resemblance on appearance between  
 the expression of solution to
  the functional equation (\ref{master}) and 
 that to the T-system,
  the solution of the functional equation (\ref{master})
 is quite different from that of the T-system.
As is often the case in soliton theory, most of the proofs of the 
determinant or pfaffian formulae reduce to the Jacobi identity. 

The outline of this paper is given as follows.
In section2, we present explicitly our functional 
equations for $A_{n}^{(1)}$, 
$B_{n}^{(1)}$, $C_{n}^{(1)}$, $D_{n}^{(1)}$, $A_{n}^{(2)}$ and 
$D_{n+1}^{(2)}$, whose solutions are given in section4.
In section3, we fix the notation.
Section5 is dedicated for summary and discussion.
 Appendix{\bf A} is a list of functional equations for $G_{2}^{(1)}$,
 $F_{4}^{(1)}$, $E_{6}^{(1)}$, $E_{7}^{(1)}$, $E_{8}^{(1)}$, 
 $E_{6}^{(2)}$ and $D_{4}^{(3)}$, which have been not yet solved.

Finally, it should be noted that, for the twisted quantum affine algebras 
$U_{q}(X_{n}^{(k)})$ $(X_{n}^{(k)}=A_{n}^{(2)},D_{n}^{(2)},
E_{6}^{(2)},D_{4}^{(3)}) $,
extensions of the T-system different from our functional equations 
were proposed by Kuniba and Suzuki \cite{KS2}.
 However, they do not reduce to affine Toda field equations 
in the continuum limit.   
\eqreset
\section{Functional Equations}
For the function $T^{(a)}_m(u)$ $(m \in {\bf Z}, 
u \in {\bf C}, a\in \{0,1,\dots, r \})$, we assume 
the following boundary condition:  
 \[ T_{m}^{(a)}(u)=0 \quad {\rm for} \quad m<0, \quad T_{0}^{(a)}(u)=1. \] 
Substituting each Cartan matrix of affine Lie algebra into the 
functional equation
 (\ref{master}), we obtain explicitly the following functional equations:
\\ 
For $A_{r}^{(1)} ( r\ge 1)$, 
\begin{equation}
     T_{m}^{(a)}(u-1) T_{m}^{(a)}(u+1)  = 
    T_{m+1}^{(a)}(u) T_{m-1}^{(a)}(u)+
        T_{m}^{(a-1)}(u) T_{m}^{(a+1)}(u) \quad 0 \le a \le r
        \label{t-sysA1} 
\end{equation}
with $T_{m}^{(-1)}(u)=T_{m}^{(r)}(u)$ and  $T_{m}^{(r+1)}(u)=
T_{m}^{(0)}(u)$.\\
  \\  
For $B_{r}^{(1)} ( r\ge 3)$, 
\begin{eqnarray}
\hspace{-45pt} &&    T_{m}^{(a)}(u-1) T_{m}^{(a)}(u+1) =
        T_{m+1}^{(a)}(u) T_{m-1}^{(a)}(u)+
        T_{m}^{(2)}(u) \quad a=0, 1 
        \label{t-sysB1-1}, \\ 
\hspace{-45pt} &&    T_{m}^{(2)}(u-1) T_{m}^{(2)}(u+1)  = T_{m+1}^{(2)}(u) 
        T_{m-1}^{(2)}(u) +
        T_{m}^{(0)}(u) T_{m}^{(1)}(u) T_{m}^{(3)}(u) 
        \label{t-sysB1-2}, \\
\hspace{-45pt} &&     T_{m}^{(a)}(u-1) T_{m}^{(a)}(u+1)  =
        T_{m+1}^{(a)}(u) T_{m-1}^{(a)}(u)  
     +T_{m}^{(a-1)}(u) T_{m}^{(a+1)}(u),
      \nonumber \\ 
      \hspace{-45pt} && \hspace{200pt} 3 \le a \le r-2 ,
        \label{t-sysB1-3} \\
\hspace{-45pt} &&    T_{m}^{(r-1)}(u-1) T_{m}^{(r-1)}(u+1) = T_{m+1}^{(r-1)}(u) 
        T_{m-1}^{(r-1)}(u)+
        T_{m}^{(r-2)}(u) T_{2m}^{(r)}(u) 
        \label{t-sysB1-4}, \\
\hspace{-45pt} &&    T_{2m}^{(r)}(u-1/2) T_{2m}^{(r)}(u+1/2)  =
        T_{2m+1}^{(r)}(u) T_{2m-1}^{(r)}(u) \nonumber \\ 
\hspace{-45pt} &&  \hspace{145pt}      +
        T_{m}^{(r-1)}(u-1/2) T_{m}^{(r-1)}(u+1/2) ,
        \label{t-sysB1-5} \\ 
\hspace{-45pt} &&   T_{2m+1}^{(r)}(u-1/2) T_{2m+1}^{(r)}(u+1/2)  =  
        T_{2m+2}^{(r)}(u) T_{2m}^{(r)}(u) \nonumber \\ 
\hspace{-45pt} &&  \hspace{160pt} +
        T_{m}^{(r-1)}(u) T_{m+1}^{(r-1)}(u) .
        \label{t-sysB1-6}
\end{eqnarray} 
If $r=3$, eqs. (\ref{t-sysB1-2}) - (\ref{t-sysB1-4}) 
reduce to the following equation :
\begin{equation}
T_{m}^{(2)}(u-1) T_{m}^{(2)}(u+1)  =  T_{m+1}^{(2)}(u) 
        T_{m-1}^{(2)}(u)+
        T_{m}^{(0)}(u) T_{m}^{(1)}(u) T_{2m}^{(3)}(u) 
        \label{t-sysB1-7} .
\end{equation}
 \quad  \\     
 For $C_{r}^{(1)} ( r\ge 2)$, 
\begin{eqnarray}
\hspace{-45pt} &&    T_{m}^{(0)}(u-1) T_{m}^{(0)}(u+1) =
        T_{m+1}^{(0)}(u) T_{m-1}^{(0)}(u)+
        T_{2m}^{(1)}(u), 
        \label{t-sysC1-1} \\ 
\hspace{-45pt} &&      T_{2m}^{(1)}(u-1/2) T_{2m}^{(1)}(u+1/2)     
     = T_{2m+1}^{(1)}(u) T_{2m-1}^{(1)}(u)  \nonumber \\
\hspace{-45pt} && \hspace{145pt}
        + T_{m}^{(0)}(u-1/2) T_{m}^{(0)}(u+1/2)T_{2m}^{(2)}(u), 
        \label{t-sysC1-2} \\ 
\hspace{-45pt} &&    T_{2m+1}^{(1)}(u-1/2) T_{2m+1}^{(1)}(u+1/2) = 
        T_{2m+2}^{(1)}(u) T_{2m}^{(1)}(u) \nonumber \\ 
\hspace{-45pt} &&  \hspace{165pt}   +
        T_{m}^{(0)}(u) T_{m+1}^{(0)}(u)T_{2m+1}^{(2)}(u) , 
        \label{t-sysC1-3} \\ 
\hspace{-45pt} &&     T_{m}^{(a)}(u-1/2) T_{m}^{(a)}(u+1/2)  = 
        T_{m+1}^{(a)}(u) T_{m-1}^{(a)}(u)+
        T_{m}^{(a-1)}(u) T_{m}^{(a+1)}(u) \nonumber \\ 
\hspace{-45pt} && \hspace{220pt} 2 \le a \le r-2, 
        \label{t-sysC1-4} \\
\hspace{-45pt} &&    T_{2m}^{(r-1)}(u-1/2) T_{2m}^{(r-1)}(u+1/2) 
  = T_{2m+1}^{(r-1)}(u) T_{2m-1}^{(r-1)}(u)\nonumber \\ 
\hspace{-45pt} && \hspace{145pt}  +
        T_{2m}^{(r-2)}(u)T_{m}^{(r)}(u-1/2) T_{m}^{(r)}(u+1/2) ,
        \label{t-sysC1-5} \\ 
\hspace{-45pt} &&    T_{2m+1}^{(r-1)}(u-1/2) T_{2m+1}^{(r-1)}(u+1/2)  = 
        T_{2m+2}^{(r-1)}(u) T_{2m}^{(r-1)}(u) \nonumber \\ 
\hspace{-45pt} &&  \hspace{170pt} +
        T_{2m+1}^{(r-2)}(u)T_{m}^{(r)}(u) T_{m+1}^{(r)}(u), 
        \label{t-sysC1-6} \\ 
\hspace{-45pt} &&    T_{m}^{(r)}(u-1) T_{m}^{(r)}(u+1)  =
        T_{m+1}^{(r)}(u) T_{m-1}^{(r)}(u)+
        T_{2m}^{(r-1)}(u) .  
        \label{t-sysC1-7}
\end{eqnarray} 
If $r=2$, eqs. (\ref{t-sysC1-2}) - (\ref{t-sysC1-6})   
reduce to the following equations :
\begin{eqnarray}
\hspace{-45pt} && T_{2m}^{(1)}(u-1/2) T_{2m}^{(1)}(u+1/2)  
   = T_{2m+1}^{(1)}(u) T_{2m-1}^{(1)}(u) \nonumber \\
\hspace{-45pt}&& \hspace{50pt} +
    T_{m}^{(0)}(u-1/2) T_{m}^{(0)}(u+1/2)T_{m}^{(2)}(u-1/2) 
    T_{m}^{(2)}(u+1/2) , 
        \label{t-sysC1-8} \\ 
\hspace{-45pt}&&  T_{2m+1}^{(1)}(u-1/2) T_{2m+1}^{(1)}(u+1/2)  
  = T_{2m+2}^{(1)}(u) T_{2m}^{(1)}(u)  \nonumber \\ 
\hspace{-45pt}  && \hspace{170pt}+
        T_{m}^{(0)}(u) T_{m+1}^{(0)}(u) T_{m}^{(2)}(u) T_{m+1}^{(2)}(u) 
        \label{t-sysC1-9} .
\end{eqnarray}
  \\ 
 For $D_{r}^{(1)} ( r\ge 4)$, 
\begin{eqnarray}
\hspace{-50pt} && T_{m}^{(a)}(u-1) T_{m}^{(a)}(u+1)  =  
        T_{m+1}^{(a)}(u) T_{m-1}^{(a)}(u)+
        T_{m}^{(2)}(u) \qquad a=0, 1 ,
        \label{t-sysD1-1} \\ 
\hspace{-50pt} &&  T_{m}^{(2)}(u-1) T_{m}^{(2)}(u+1)  =  T_{m+1}^{(2)}(u) 
        T_{m-1}^{(2)}(u)+
        T_{m}^{(0)}(u) T_{m}^{(1)}(u) T_{m}^{(3)}(u) ,
        \label{t-sysD1-2} \\
\hspace{-50pt} &&   T_{m}^{(a)}(u-1) T_{m}^{(a)}(u+1)  = 
        T_{m+1}^{(a)}(u) T_{m-1}^{(a)}(u)+
        T_{m}^{(a-1)}(u) T_{m}^{(a+1)}(u) \label{t-sysD1-3} \\ 
 \hspace{-50pt} && \hspace{180pt}  3 \le a \le r-3 , \nonumber 
         \\
\hspace{-50pt} &&  T_{m}^{(r-2)}(u-1) T_{m}^{(r-2)}(u+1)  =  T_{m+1}^{(r-2)}(u) 
        T_{m-1}^{(r-2)}(u) \nonumber \\ 
\hspace{-50pt} && \hspace{160pt}      +
        T_{m}^{(r-3)}(u) T_{m}^{(r-1)}(u)T_{m}^{(r)}(u),  
        \label{t-sysD1-4} \\
\hspace{-50pt} && T_{m}^{(a)}(u-1) T_{m}^{(a)}(u+1)  =  
        T_{m+1}^{(a)}(u) T_{m-1}^{(a)}(u)+
        T_{m}^{(r-2)}(u) \ a=r-1, r.  
        \label{t-sysD1-5} 
\end{eqnarray} 
If $r=4$, eqs. (\ref{t-sysD1-2}) - (\ref{t-sysD1-4}) 
reduce to the following equation :
\begin{equation}
T_{m}^{(2)}(u-1) T_{m}^{(2)}(u+1)  =  T_{m+1}^{(2)}(u) 
        T_{m-1}^{(2)}(u)+
        T_{m}^{(0)}(u) T_{m}^{(1)}(u)T_{m}^{(r-1)}(u) T_{m}^{(r)}(u) 
        \label{t-sysD1-6} .
\end{equation}
\quad  \\    
For $A_{2r}^{(2)} ( r\ge 2)$, 
\begin{eqnarray}
\hspace{-45pt} &&     T_{m}^{(0)}(u-1) T_{m}^{(0)}(u+1)  =  
        T_{m+1}^{(0)}(u) T_{m-1}^{(0)}(u)+
        T_{2m}^{(1)}(u), 
        \label{t-sysA2-1} \\ 
\hspace{-45pt} &&       T_{2m}^{(1)}(u-1/2) T_{2m}^{(1)}(u+1/2)  \nonumber \\   
\hspace{-45pt} &&      = T_{2m+1}^{(1)}(u) T_{2m-1}^{(1)}(u)  
        + T_{m}^{(0)}(u-1/2) T_{m}^{(0)}(u+1/2)T_{2m}^{(2)}(u) ,
        \label{t-sysA2-2} \\ 
\hspace{-45pt} &&     T_{2m+1}^{(1)}(u-1/2) T_{2m+1}^{(1)}(u+1/2)  =  
        T_{2m+2}^{(1)}(u) T_{2m}^{(1)}(u) \nonumber \\ 
\hspace{-45pt} && \hspace{165pt}
+T_{m}^{(0)}(u) T_{m+1}^{(0)}(u)T_{2m+1}^{(2)}(u), 
        \label{t-sysA2-3} \\ 
\hspace{-45pt} &&     T_{m}^{(a)}(u-1/2) T_{m}^{(a)}(u+1/2)  = 
        T_{m+1}^{(a)}(u) T_{m-1}^{(a)}(u) \label{t-sysA2-4} \\
 \hspace{-45pt}     &&   \hspace{130pt}       
 +T_{m}^{(a-1)}(u) T_{m}^{(a+1)}(u)  \quad 
 2 \le a \le r-2 ,
         \nonumber \\
\hspace{-45pt} &&     T_{m}^{(r-1)}(u-1/2) T_{m}^{(r-1)}(u+1/2) 
        =  T_{m+1}^{(r-1)}(u)T_{m-1}^{(r-1)}(u)\nonumber \\
 \hspace{-45pt} && \hspace{170pt}+
        T_{m}^{(r-2)}(u) T_{2m}^{(r)}(u), 
        \label{t-sysA2-5} \\
\hspace{-45pt} &&     T_{2m}^{(r)}(u-1/4) T_{2m}^{(r)}(u+1/4)  =  
        T_{2m+1}^{(r)}(u) T_{2m-1}^{(r)}(u) \nonumber \\ 
\hspace{-45pt} && \hspace{160pt} + 
        T_{m}^{(r-1)}(u-1/4) T_{m}^{(r-1)}(u+1/4) ,
        \label{t-sysA2-6} \\ 
\hspace{-45pt} &&         T_{2m+1}^{(r)}(u-1/4) T_{2m+1}^{(r)}(u+1/4)  =  
        T_{2m+2}^{(r)}(u) T_{2m}^{(r)}(u)\nonumber \\
 \hspace{-45pt} && \hspace{170pt}  +
        T_{m}^{(r-1)}(u) T_{m+1}^{(r-1)}(u) .
        \label{t-sysA2-7}
\end{eqnarray} 
If $r=2$, eqs. (\ref{t-sysA2-2}) - (\ref{t-sysA2-5})  
reduce to the following equations :
\begin{eqnarray}
\hspace{-45pt} &&    T_{2m}^{(1)}(u-1/2) T_{2m}^{(1)}(u+1/2) \nonumber \\ 
\hspace{-45pt} &&    = T_{2m+1}^{(1)}(u) T_{2m-1}^{(1)}(u)+
        T_{m}^{(0)}(u-1/2) T_{m}^{(0)}(u+1/2)T_{4m}^{(2)}(u) ,
        \label{t-sysA2-8} \\ 
\hspace{-45pt} &&    T_{2m+1}^{(1)}(u-1/2) T_{2m+1}^{(1)}(u+1/2) 
     = T_{2m+2}^{(1)}(u) T_{2m}^{(1)}(u) \nonumber \\ 
 \hspace{-45pt} &&\hspace{165pt}  +
        T_{m}^{(0)}(u) T_{m+1}^{(0)}(u) T_{4m+2}^{(2)}(u) 
        \label{t-sysA2-9} .
\end{eqnarray} 
  \\    
 For $A_{2r-1}^{(2)} ( r\ge 3)$, 
\begin{eqnarray}
\hspace{-60pt} &&    T_{m}^{(a)}(u-1/2) T_{m}^{(a)}(u+1/2)  =  
        T_{m+1}^{(a)}(u) T_{m-1}^{(a)}(u)+
        T_{m}^{(2)}(u) \ a=0, 1 ,
        \label{t-sysA22-1} \\ 
\hspace{-60pt} &&    T_{m}^{(2)}(u-1/2) T_{m}^{(2)}(u+1/2)  =  T_{m+1}^{(2)}(u) 
        T_{m-1}^{(2)}(u) \nonumber \\
        \hspace{-45pt} &&\hspace{155pt} +
        T_{m}^{(0)}(u) T_{m}^{(1)}(u) T_{m}^{(3)}(u), 
        \label{t-sysA22-2} \\
\hspace{-60pt} &&     T_{m}^{(a)}(u-1/2) T_{m}^{(a)}(u+1/2)  = 
        T_{m+1}^{(a)}(u) T_{m-1}^{(a)}(u)\label{t-sysA22-3}\\
 \hspace{-60pt} &&  \hspace{150pt}       
 +T_{m}^{(a-1)}(u) T_{m}^{(a+1)}(u) \quad  
    3 \le a \le r-2 \nonumber \\
\hspace{-60pt} &&    T_{2m}^{(r-1)}(u-1/2) T_{2m}^{(r-1)}(u+1/2) \nonumber \\
\hspace{-60pt} &&  = T_{2m+1}^{(r-1)}(u) T_{2m-1}^{(r-1)}(u)+
        T_{2m}^{(r-2)}(u)T_{m}^{(r)}(u-1/2) T_{m}^{(r)}(u+1/2), 
        \label{t-sysA22-4} \\ 
\hspace{-60pt} &&    T_{2m+1}^{(r-1)}(u-1/2) T_{2m+1}^{(r-1)}(u+1/2)  =  
        T_{2m+2}^{(r-1)}(u) T_{2m}^{(r-1)}(u) \nonumber \\
\hspace{-60pt} && \hspace{170pt} +
        T_{2m+1}^{(r-2)}(u)T_{m}^{(r)}(u) T_{m+1}^{(r)}(u) ,
        \label{t-sysA22-5} \\ 
\hspace{-60pt} &&    T_{m}^{(r)}(u-1) T_{m}^{(r)}(u+1)  =  
        T_{m+1}^{(r)}(u) T_{m-1}^{(r)}(u)+
        T_{2m}^{(r-1)}(u).  
        \label{t-sysA22-6}
\end{eqnarray} 
If $r=3$, eqs. (\ref{t-sysA22-2}) - (\ref{t-sysA22-5}) 
reduce to the following equations :
\begin{eqnarray}
\hspace{-45pt} &&    T_{2m}^{(2)}(u-1/2) T_{2m}^{(2)}(u+1/2) \nonumber \\
\hspace{-45pt} &&  = T_{2m+1}^{(2)}(u) T_{2m-1}^{(2)}(u)+
        T_{2m}^{(0)}(u)T_{2m}^{(1)}(u)T_{m}^{(3)}(u-1/2)
        T_{m}^{(3)}(u+1/2),
        \label{t-sysA22-7} \\ 
\hspace{-45pt} &&    T_{2m+1}^{(2)}(u-1/2) T_{2m+1}^{(2)}(u+1/2)  \nonumber
\\   
\hspace{-45pt} &&      = T_{2m+2}^{(2)}(u) T_{2m}^{(2)}(u)+
        T_{2m+1}^{(0)}(u)T_{2m+1}^{(1)}(u)T_{m}^{(3)}(u) T_{m+1}^{(3)}(u). 
        \label{t-sysA22-8} 
\end{eqnarray}
 \\ 
 For $D_{r+1}^{(2)} ( r\ge 2)$, 
\begin{eqnarray}
\hspace{-45pt} &&    T_{2m}^{(0)}(u-1/2) T_{2m}^{(0)}(u+1/2)  =  
        T_{2m+1}^{(0)}(u) T_{2m-1}^{(0)}(u)\nonumber \\ 
\hspace{-45pt} &&\hspace{165pt}   +
        T_{m}^{(1)}(u-1/2) T_{m}^{(1)}(u+1/2) ,
        \label{t-sysD2-1} \\ 
\hspace{-45pt} &&    T_{2m+1}^{(0)}(u-1/2) T_{2m+1}^{(0)}(u+1/2)  =  
        T_{2m+2}^{(0)}(u) T_{2m}^{(0)}(u)\nonumber \\
\hspace{-45pt} &&\hspace{170pt} +
        T_{m}^{(1)}(u) T_{m+1}^{(1)}(u), 
        \label{t-sysD2-2} \\
\hspace{-45pt} &&    T_{m}^{(1)}(u-1) T_{m}^{(1)}(u+1)  =  T_{m+1}^{(1)}(u) 
        T_{m-1}^{(1)}(u)+
        T_{2m}^{(0)}(u) T_{m}^{(2)}(u), 
        \label{t-sysD2-3} \\
\hspace{-45pt} &&     T_{m}^{(a)}(u-1) T_{m}^{(a)}(u+1)  = 
        T_{m+1}^{(a)}(u) T_{m-1}^{(a)}(u)+
        T_{m}^{(a-1)}(u) T_{m}^{(a+1)}(u) \label{t-sysD2-4} \\ 
\hspace{-45pt} && \hspace{200pt}   2 \le a \le r-2, \nonumber 
         \\
\hspace{-45pt} &&    T_{m}^{(r-1)}(u-1) T_{m}^{(r-1)}(u+1)  = 
T_{m+1}^{(r-1)}(u) 
        T_{m-1}^{(r-1)}(u)\nonumber \\
\hspace{-45pt} && \hspace{170pt}  +
        T_{m}^{(r-2)}(u) T_{2m}^{(r)}(u), 
        \label{t-sysD2-5} \\
\hspace{-45pt} &&    T_{2m}^{(r)}(u-1/2) T_{2m}^{(r)}(u+1/2) \nonumber \\    
\hspace{-45pt} &&  = T_{2m+1}^{(r)}(u) T_{2m-1}^{(r)}(u)+
        T_{m}^{(r-1)}(u-1/2) T_{m}^{(r-1)}(u+1/2) ,
        \label{t-sysD2-6} \\ 
\hspace{-45pt} &&        T_{2m+1}^{(r)}(u-1/2) T_{2m+1}^{(r)}(u+1/2)  =  
        T_{2m+2}^{(r)}(u) T_{2m}^{(r)}(u) \nonumber \\ 
\hspace{-45pt} &&  \hspace{170pt} +
        T_{m}^{(r-1)}(u) T_{m+1}^{(r-1)}(u) .
        \label{t-sysD2-7}
\end{eqnarray} 
If $r=2$, eqs. (\ref{t-sysD2-3}) - (\ref{t-sysD2-5}) 
reduce to the following equation :
\begin{equation}
    T_{m}^{(1)}(u-1) T_{m}^{(1)}(u+1)  =  T_{m+1}^{(1)}(u) 
        T_{m-1}^{(1)}(u)+
         T_{2m}^{(0)}(u)T_{2m}^{(2)}(u). 
        \label{t-sysD2-8} 
\end{equation}
These functional equations are neatly connected with Dynkin diagrams, 
and thus they are reflecting Dynkin diagram symmetries.
For example, the functional equations (\ref{t-sysD1-1}) and 
(\ref{t-sysD1-2}) are transformed into eqs.(\ref{t-sysD1-5}) 
and (\ref{t-sysD1-4}) respectively 
under the following transformation:
\begin{equation}
  \quad T_{m}^{(b)}(u) \to T_{m}^{(r-b)}(u).
\end{equation}
It was pointed out \cite{KNH} that the eq.(\ref{t-sysA1}) 
with boundary condition 
\[ T_{m}^{(-1)}(u)=0 ,\quad  T_{m}^{(0)}(u)=T_{m}^{(r+1)}(u)=1 \]
corresponds to Hirota-Miwa equation.

 Kuniba {\it  et al.} \cite{KOS} pointed 
 out that the T-system may be 
looked upon as 
a discritized Toda field equation. By the same argument, our functional 
equation (\ref{master}) may be viewed as a discritized 
affine Toda field equation.
Set $u=v/\delta$ and $m=w/\delta$, 
for functional equation (\ref{master}).
For small $\delta $,
 from the coefficient of $\delta^2$ on Taylor expansion, we obtain  
 two-dimensional affine Toda field equation of the form 
\begin{equation}
(\partial_{v}^{2}-\partial_{w}^{2}) \log \psi_{a}(v,w)=constant 
\prod_{b=0}^{r} \psi_{b}(v,w)^{-C_{a b}}
\end{equation} 
where $\psi_{a}(v,w)$ denotes a scaled $T_{m}^{(a)}(u)$ and 
$C_{a b}=2(\alpha_{a} | \alpha_{b})/(\alpha_{a} | \alpha_{a})$
 the Cartan matrix of affine Lie algebra.
\eqreset
\section{Notation and Convention}
In this section, we present a lot of expressions, which are necessary 
to construct the solutions. The origin of them go back to earlier 
work \cite{KOS,KNH,TK} on solutions of the T-system.  
Set  
\begin{eqnarray}
\hspace{-10pt} &&
 x_{j}^{[1]}(k|u)=T_{1}^{(\delta_{j-1})}(u+(2j-2)/k), \quad 
y_{j}^{[1]}(k|u)=T_{1}^{(\delta_{j})}(u+(2j-2)/k) ,\nonumber \\   
\hspace{-10pt} &&
t_{ij}^{[1]}(k|u)={\cal T}^{1-i+j}(u+(i+j-2)/k),\nonumber \\ 
\hspace{-10pt} && x_{j}^{[2]}(k|u)=T_{1}^{(r-\delta_{j-1})}(u+(2j-2)/k), \quad 
y_{j}^{[2]}(k|u)=T_{1}^{(r-\delta_{j})}(u+(2j-2)/k) ,\nonumber \\   
\hspace{-10pt} &&
t_{ij}^{[2]}(k|u)={\cal T}^{r+i-j-1}(u+(i+j-2)/k),  \\
\hspace{-10pt} && a_{ij}^{[p]}(k|u)=
x_{i}^{[p]}(k|u)y_{j}^{[p]}(k|u)-t_{ij}^{[p]}(k|u), \nonumber \\ 
\hspace{-10pt} && 
b_{ij}^{[p]}(k|u)=y_{i}^{[p]}(k|u)x_{j}^{[p]}(k|u)-t_{ij}^{[p]}(k|u), 
\; p=1,2. \nonumber 
\end{eqnarray}
where 
\[ \delta_{i}= 
     \left\{
            \begin{array}{ll}
            0 & {\rm if} \quad i \in 2{\bf Z } \\
            1 & {\rm if} \quad i \in 2{\bf Z }+1 . 
            \end{array}
    \right.
 \]
By the definition we have  
\begin{eqnarray}
\hspace{-40pt} &&    x_{i}^{[p]}(k|u+2/k)=y_{i+1}^{[p]}(k|u), 
         \quad y_{i}^{[p]}(k|u+2/k)=x_{i+1}^{[p]}(k|u),\nonumber \\   
\hspace{-40pt} &&    t_{ij}^{[p]}(k|u+2/k)=t_{i+1 \; j+1}^{[p]}(k|u),
\nonumber \\
\hspace{-40pt} &&  a_{ij}^{[p]}(k|u+2/k)=b_{i+1 \; j+1}^{[p]}(k|u), \quad 
        b_{ij}^{[p]}(k|u+2/k)=a_{i+1 \; j+1}^{[p]}(k|u) , 
        \label{relation-sift}
\end{eqnarray}
for $p=1,2$. 
These relations are necessary to prove Lemma\ref{lemma2}.
Now we introduce $m\times m$ matrices 
${\cal T}_{m}^{a}(k|u) =({\cal T}_{ij}^{a}(k|u))_{1\le i,j\le m}$ 
$(a \in {\bf Z})$ , $(2m)\times (2m)$ matrices  
${\cal F}_{2m}^{[p]}(k|u)=({\cal F}_{ij}^{[p]}(k|u))_{1\le i,j\le 2m}$ 
, $(m+1)\times (m+1)$ matrices  
${\cal H}_{m+1}^{[p]}(k|u)=({\cal H}_{ij}^{[p]}(k|u))_{1\le i,j\le m+1}$,
${\cal S}_{m+1}^{[p]}(k|u)=({\cal S}_{ij}^{[p]}(k|u))_{1\le i,j\le m+1}$,
 $ {\cal C}_{m+1}^{[p]}(k|u)=
({\cal C}_{ij}^{[p]}(k|u))_{1\le i,j\le m+1} $ and $ 
{\cal R}_{m+1}^{[p]}(k|u)=({\cal R}_{ij}^{[p]}(k|u))_{1\le i,j\le m+1}$
 $(p=1,2 )$
 whose $(i,j)$ elements are given by 
\begin{equation}
 {\cal T}_{ij}^{a}(k|u) =
            {\cal T}^{a+i-j}(u+(i+j-m-1)/k),  
\end{equation} 
\begin{eqnarray}
\hspace{-45pt} && {\cal F}_{i j}^{[1]}(k|u) =
  \left\{
          \begin{array}{ll}
            0 & 
            {\rm : } \quad i=j \\
            {\cal T}^{-i+j-1}(u+(i+j-2m-1)/k) & 
            {\rm : } \quad 1 \le i < j \le 2m \\ 
            -{\cal T}^{i-j-1}(u+(i+j-2m-1)/k) & 
            {\rm : } \quad 1 \le j < i \le 2m ,
      \end{array}
    \right. \\ 
\hspace{-45pt} &&{\cal F}_{i j}^{[2]}(k|u) =
  \left\{
          \begin{array}{ll}
            0 & {\rm : } \quad i=j \\
            {\cal T}^{i-j+r+1}(u+(i+j-2m-1)/k) & \\ 
    \hspace{90pt} {\rm : } \quad 1 \le i < j \le 2m & \\ 
            -{\cal T}^{-i+j+r+1}(u+(i+j-2m-1)/k) & 
            {\rm : } \quad 1 \le j < i \le 2m ,
      \end{array}
    \right.
\end{eqnarray}
\begin{eqnarray}
 {\cal H}_{ij}^{[1]}(k|u) =
    \left\{
          \begin{array}{ll}
            T_{1}^{(0)}(u+(2i-2)/k)  \quad 
            {\rm : }  1 \le i \le m+1 \ , {\rm and} \ , j=1& \\
            {\cal T}^{-i+j}(u+(i+j-5/2)/k)  & \\ 
             \hspace{70pt} {\rm : } \quad 1 \le i \le m+1, 
            \quad 2 \le j \le m+1 , &
      \end{array}
    \right.
\end{eqnarray}
\begin{eqnarray}
 {\cal H}_{ij}^{[2]}(k|u) =
    \left\{
          \begin{array}{ll}
            T_{1}^{(r)}(u+(2i-2)/k) \quad  
            {\rm : }  1 \le i \le m+1 \, {\rm and} \, j=1 & \\
            {\cal T}^{r+i-j}(u+(i+j-5/2)/k) & \\ 
   \hspace{70pt} {\rm : } 1 \le i \le m+1, 
            \quad 2 \le j \le m+1 , &
      \end{array}
    \right.
\end{eqnarray}
\begin{eqnarray}
 {\cal S}_{ij}^{[1]}(k|u) =
    \left\{
          \begin{array}{ll}
            0 & 
            {\rm : }  i=j=1 \\
            T_{1}^{(\delta_{j})}(u+(2j-4)/k) & 
            {\rm : }   i=1 \, {\rm and} \, 2 \le j \le m+1 \\
            -T_{1}^{(\delta_{i})}(u+(2i-4)/k) & 
            {\rm : }  2 \le i \le m+1 \, {\rm and} \, j=1 \\
            -{\cal T}^{1-i+j}(u+(i+j-4)/k) & 
            {\rm : }  2 \le i, j \le m+1,
      \end{array}
    \right.
\end{eqnarray}
\begin{eqnarray}
 {\cal S}_{ij}^{[2]}(k|u) =
    \left\{
          \begin{array}{ll}
            0 & 
            {\rm : }  i=j=1 \\
            T_{1}^{(r-\delta_{j})}(u+(2j-4)/k) & \\ 
 \hspace{60pt} {\rm : } i=1 \, {\rm and} \, 2 \le j \le m+1 & \\
            -T_{1}^{(r-\delta_{i})}(u+(2i-4)/k) & \\  
 \hspace{70pt} {\rm : }  2 \le i \le m+1 \, and \, j=1  & \\
            -{\cal T}^{r+i-j-1}(u+(i+j-4)/k) & 
            {\rm : }  2 \le i, j \le m+1,
      \end{array}
    \right.
\end{eqnarray}
\begin{eqnarray}
\hspace{-45pt} &&  {\cal C}_{i j}^{[p]}(k|u) =
  \left\{
          \begin{array}{ll}
            0 & 
            {\rm : } \quad i=j \\
            x_{j-1}^{[p]}(k|u) & 
            {\rm : } \quad  i=1 \quad and \quad 2 \le j \le m+1 \\
            -x_{i-1}^{[p]}(k|u) & 
            {\rm : } \quad 2 \le i \le m+1 \quad and \quad j=1 \\
            a_{i-1 \; j-1}^{[p]}(k|u) & 
            {\rm : } \quad 2 \le i < j \le m+1 \\ 
            -a_{j-1 \; i-1}^{[p]}(k|u) & 
            {\rm : } \quad 2 \le j < i \le m+1,
      \end{array}
    \right. \\ 
\hspace{-45pt} &&  {\cal R}_{i j}^{[p]}(k|u) =
  \left\{
          \begin{array}{ll}
            0 & 
            {\rm : } \quad i=j \\
            x_{j-1}^{[p]}(k|u) & 
            {\rm : } \quad  i=1 \quad and \quad 2 \le j \le m+1 \\
            -x_{i-1}^{[p]}(k|u) & 
            {\rm : } \quad 2 \le i \le m+1 \quad and \quad j=1 \\
            b_{i-1 \; j-1}^{[p]}(k|u) & 
            {\rm : } \quad 2 \le i < j \le m+1 \\ 
            -b_{j-1 \; i-1}^{[p]}(k|u) & 
            {\rm : } \quad 2 \le j < i \le m+1.
      \end{array}
    \right.
\end{eqnarray} 
Note that the matrices ${\cal F}_{i j}^{[1]}(k|u)$, 
${\cal H}_{ij}^{[1]}(k|u)$, ${\cal S}_{ij}^{[1]}(k|u)$,   
${\cal C}_{i j}^{[1]}(k|u)$ and ${\cal R}_{i j}^{[1]}(k|u)$ 
 are transformed into ${\cal F}_{i j}^{[2]}(k|u)$, 
${\cal H}_{ij}^{[2]}(k|u)$, ${\cal S}_{ij}^{[2]}(k|u)$,   
${\cal C}_{i j}^{[2]}(k|u)$ and ${\cal R}_{i j}^{[2]}(k|u)$ 
respectively under the following transformation:
\begin{equation}
   T_{1}^{(b)}(u) \to T_{1}^{(r-b)}(u), \qquad 
 {\cal T}^{b}(u) \to {\cal T}^{r-b}(u). \label{trans}
\end{equation}
For any matrix ${\cal M}(u)$, we shall let 
${\cal M}\left[
    \begin{array}{ccc}
          i_{1} & \dots & i_{k}  \\
          j_{1} & \dots & j_{k}
    \end{array}
  \right](u)$
denote the minor matrix  getting rid of  
$i_{l} $\symbol{39}s rows and $j_{l} 
$\symbol{39}s columns from ${\cal M}(u)$.
We introduce the following pfaffian and determinant   
expressions, which will be used as the components of the solutions.   
\begin{eqnarray}
\hspace{-45pt} &&   T_{m}^{a}(k|u)=
       \det [{\cal T}_{m}^{a}(k|u)]
       \quad {\rm for} 
       \ m \in {\bf Z }_{\ge 0},
      \label{Ta} \\
\hspace{-45pt} &&   F_{m}^{(0)}(k|u)=
       {\rm pf} [{\cal F}_{2m}^{[1]}(k|u)]
       \quad {\rm for} 
       \ m \in {\bf Z }_{\ge 0},
      \label{F0} \\
\hspace{-45pt} &&   F_{m}^{(r)}(k|u)=
       {\rm pf} [{\cal F}_{2m}^{[2]}(k|u)]
       \quad {\rm for} 
      \ m \in {\bf Z }_{\ge 0},
      \label{Fr} \\
\hspace{-45pt} &&    G_{m}^{(0)}(k|u)=
           \left\{
           \begin{array}{ll}
            {\rm pf}[{\cal C}_{m+1}^{[1]}
              \left[
              \begin{array}{c}
               1 \\
               1
              \end{array}
              \right]
              (k|u+(-m+1)/k)] & 
              {\rm for} \quad  m \in 2{\bf Z }_{\ge 0} \\
            {\rm pf}[{\cal C}_{m+1}^{[1]}(k|u+(-m+1)/k)] & 
             {\rm for} \quad  m \in 2{\bf Z }_{\ge 0}+1,
            \end{array}
           \right.
    \label{G0} \\
\hspace{-45pt} &&    G_{m}^{(1)}(k|u)=
           \left\{
           \begin{array}{ll}
            {\rm pf}[{\cal C}_{m+2}^{[1]}
              \left[
              \begin{array}{cc}
               1 & 2\\
               1 & 2
              \end{array}
              \right]
              (k|u+(-m-1)/k)] & 
             {\rm for} \quad  m \in 2{\bf Z }_{\ge 0} \\
            {\rm pf}[{\cal C}_{m+2}^{[1]}
            \left[
              \begin{array}{c}
               2 \\
               2
              \end{array}
              \right]
            (k|u+(-m-1)/k)] &
            {\rm for} \qquad  m \in 2{\bf Z }_{\ge 0}+1,
            \end{array}
           \right.
        \label{G1} \\ 
\hspace{-45pt} &&    G_{m}^{(r)}(k|u)=
           \left\{
           \begin{array}{ll}
            {\rm pf}[{\cal C}_{m+1}^{[2]}
              \left[
              \begin{array}{c}
               1 \\
               1
              \end{array}
              \right]
              (k|u+(-m+1)/k)] & 
              {\rm for} \quad  m \in 2{\bf Z }_{\ge 0} \\
            {\rm pf}[{\cal C}_{m+1}^{[2]}(k|u+(-m+1)/k)] & 
             {\rm for} \quad  m \in 2{\bf Z }_{\ge 0}+1,
            \end{array}
           \right.
        \label{Gr} \\
\hspace{-45pt} &&    G_{m}^{(r-1)}(k|u)=
           \left\{
           \begin{array}{ll}
            {\rm pf}[{\cal C}_{m+2}^{[2]}
              \left[
              \begin{array}{cc}
               1 & 2\\
               1 & 2
              \end{array}
              \right]
              (k|u+(-m-1)/k)] & 
             {\rm for} \quad  m \in 2{\bf Z }_{\ge 0} \\
            {\rm pf}[{\cal C}_{m+2}^{[2]}
            \left[
              \begin{array}{c}
               2 \\
               2
              \end{array}
              \right]
            (k|u+(-m-1)/k)] &
            {\rm for} \qquad  m \in 2{\bf Z }_{\ge 0}+1,
            \end{array}
           \right.
        \label{Gr-1}   \\ 
\hspace{-45pt} &&  H_{m+1}^{(0)}(k|u)=\det [{\cal H}_{m+1}^{[1]}(k|u-m/k)] 
  \quad {\rm for} \quad  m \in {\bf Z }_{\ge 0}, \label{H0}\\
\hspace{-45pt} &&  H_{m+1}^{(r)}(k|u)=\det [{\cal H}_{m+1}^{[2]}(k|u-m/k)]
  \quad {\rm for} \quad  m \in {\bf Z }_{\ge 0}.  \label{Hr}  
\end{eqnarray}
For $a \in {\bf Z}$, $k \in {\bf Z-\{0\}}$ and $u \in {\bf C}$, 
we introduce the following functional relations, 
combinations of which will be used in solving the eq.(\ref{master}). 
\begin{equation}
{\cal T}^{a}(u)+{\cal T}^{1-a}(u) \label{rel1} 
 =T_{1}^{(0)}(u+(a-1/2)/k)T_{1}^{(0)}(u+(-a+1/2)/k) ,
\end{equation}
\begin{equation}
{\cal T}^{a}(u)+{\cal T}^{2r-1-a}(u) \label{rel2} 
 =T_{1}^{(r)}(u+(-r+a+1/2)/k)T_{1}^{(r)}(u+(r-a-1/2)/k) ,
\end{equation}
\begin{equation}
 {\cal T}^{a}(u)+{\cal T}^{-2-a}(u) =0, \label{rel3}
\end{equation}
\begin{equation}
 {\cal T}^{a}(u)+{\cal T}^{2r-a+2}(u) =0, \label{rel4}
\end{equation}
\begin{eqnarray}
\hspace{-45pt} &&{\cal T}^{a}(u)+{\cal T}^{2-a}(u) \label{fun.rel.} 
 =T_{1}^{(0)}(u+(a-1)/k)T_{1}^{(\delta_{a})}(u+(-a+1)/k) \nonumber  \\
\hspace{-45pt} && \hspace{55pt}
+T_{1}^{(1)}(u+(a-1)/k)T_{1}^{(\delta_{a-1})}(u+(-a+1)/k),\label{rel5} 
\end{eqnarray}
\begin{eqnarray}
\hspace{-45pt} && {\cal T}^{a}(u)+{\cal T}^{2r-a-2}(u) \label{fun.rel.} 
 =T_{1}^{(r)}(u+(r-a-1)/k)T_{1}^{(r-\delta_{r-a})}(u+(-r+a+1)/k) 
 \nonumber  \\
\hspace{-45pt} && \hspace{55pt} 
+T_{1}^{(r-1)}(u+(r-a-1)/k)T_{1}^{(r-\delta_{r-a-1})}(u+(-r+a+1)/k). 
\label{rel6}
\end{eqnarray}
Note that the functional equations (\ref{rel1}), (\ref{rel3}) and 
(\ref{rel5}) are transformed into eqs.(\ref{rel2}), (\ref{rel4})
and (\ref{rel6}) respectively 
under the following transformation (\ref{trans}).

The origin of the function ${\cal T}^{a}(u)$ in eq.(\ref{rel2}) goes
 back to auxiliary transfer matrix
  (or `dress function' in the analytic Bethe ansatz) for 
  $U_q(B_r^{(1)})$ vertex model.
Actually, the functional relation (\ref{rel2}) for $k=1$
 with the boundary condition  
\begin{equation}
 {\cal T}^a(u)= 
\left\{
     \begin{array}{ll}
      0              & a<0 \\ 
      1              & a=0 \\ 
      T_{1}^{(a)}(u) & 1 \le a \le r-1        
     \end{array}
 \right.
\end{equation}  
was firstly introduced in ref.\cite{KOS}.
The origin of the function ${\cal T}^{a}(u)$ in eq.(\ref{rel6}) goes
 back to auxiliary transfer matrix
  (or `dress function' in the analytic Bethe ansatz) for 
  $U_q(D_r^{(1)})$ vertex model.
Actually, the functional relation (\ref{rel6}) for $k=1$
 with the boundary condition 
\begin{equation}
 {\cal T}^a(u)= 
\left\{
     \begin{array}{ll}
      0              & a<0 \\ 
      1              & a=0 \\ 
      T_{1}^{(a)}(u) & 1 \le a \le r-2        
     \end{array}
 \right. \label{boun}
\end{equation}   
was firstly introduced in ref.\cite{TK}. \\  
Remark1 : Under the functional relation (\ref{rel5}) (resp., (\ref{rel6}))
, the following relations for $p=1$ (resp., $p=2$) hold.   
\begin{eqnarray}  
{\cal C}_{m+1}^{[p]}(k|u) &=&
{\cal S}_{m+1}^{[p]}(k|u) \prod_{j=2}^{m+1}
   P(1,j;-y_{j-1}^{[p]}(k|u)) ,\\ 
{\cal R}_{m+1}^{[p]}(k|u) &=&
\left( \prod_{i=2}^{m+1} P(i,1;y_{i-1}^{[p]}(k|u)) \right) 
{\cal S}_{m+1}^{[p]}(k|u)
\end{eqnarray} 
where 
\begin{equation}
P(i,j;c)=I+cI_{ij} \nonumber 
\end{equation}
is the $m+1$ by $m+1$ matrix with $I$ the identity and 
$I_{i j}$ the matrix unit.  Namely, the matrices
 ${\cal C}_{m+1}^{[p]}(k|u)$ and ${\cal R}_{m+1}^{[p]}(k|u)$ can be 
derived from elementary transformations 
to the matrix ${\cal S}_{m+1}^{[p]}(k|u)$.

We further have  
\begin{equation}
      {\cal F}_{2m}^{[1]}(k|u)=
        ^{t} \! {\cal T}_{2m}^{-1}(k|u)
       =-{\cal T}_{2m}^{-1}(k|u) 
\end{equation} 
 under the functional relations (\ref{rel3}),  
\begin{equation}
        {\cal F}_{2m}^{[2]}(k|u)=
        {\cal T}_{2m}^{r+1}(k|u)
\end{equation}  
under the functional relations (\ref{rel4}), 
where the index $^t$ denotes transposition of a matrix. \\ 
Remark2 : In deriving the solutions in section4, the parameter $k$ in 
this section is determined by the value of 
$t_a=2/(\alpha_{a}|\alpha_{a})$, where $t_a=1$ if $\alpha_{a}$ 
is the longest simple root of affine Lie algebra.
\eqreset
\section{Solutions}
The solutions of our functional equations 
(\ref{t-sysA1})-(\ref{t-sysD2-8}) are given as follows.
\begin{theorem}({\it The $A_{r}^{(1)}$ case, $r \ge 1$})
For $m\in Z_{\ge 0}$, 
\begin{equation}
T_{m}^{(a)}(u)= T_{m}^{a}(1|u) 	\label{sol-A1}
\end{equation}
solves the functional equations (\ref{t-sysA1}) under the condition 
\begin{equation}
 {\cal T}^{b+g}(u)={\cal T}^{b}(u),
\quad g=r+1, \quad b \in {\bf Z}. 	
\end{equation} 
\end{theorem}
\begin{theorem}({\it The $B_{r}^{(1)}$ case, $r \ge 3$})
For $m\in Z_{\ge 0}$, 
\begin{eqnarray}
T_{m}^{(0)}(u)&=&G_{m}^{(0)}(1|u), \quad   
T_{m}^{(1)}(u)=G_{m}^{(1)}(1|u), \nonumber \\ 
T_{m}^{(a)}(u)&=&T_{m}^{a}(1|u) ,\quad 2 \le a \le r-1, \label{sol-B1} \\
T_{2m}^{(r)}(u)&=&T_{m}^{r}(1|u) , \quad 
T_{2m+1}^{(r)}(u)=H_{m+1}^{(r)}(1|u) \nonumber 	
\end{eqnarray} 
solves the functional equations (\ref{t-sysB1-1}-\ref{t-sysB1-7}) under 
the relations (\ref{rel2}) and (\ref{rel5}) for $k=1$.
\end{theorem}
\begin{theorem}({\it The $C_{r}^{(1)}$ case, $r \ge 2$})
For $m\in Z_{\ge 0}$, 
\begin{eqnarray}
T_{m}^{(0)}(u)&=&F_{m}^{(0)}(2|u), \nonumber \\  
T_{m}^{(a)}(u)&=&T_{m}^{a}(2|u) ,\quad 1 \le a \le r-1 \label{sol-C1} \\
T_{m}^{(r)}(u)&=&F_{m}^{(r)}(2|u) \nonumber 	
\end{eqnarray} 
solves the functional equations (\ref{t-sysC1-1}-\ref{t-sysC1-9})
 under the relations (\ref{rel3}) and (\ref{rel4}).
\end{theorem}
\begin{theorem}({\it The $D_{r}^{(1)}$ case, $r \ge 4$})
For $m\in Z_{\ge 0}$, 
\begin{eqnarray}
T_{m}^{(0)}(u)&=&G_{m}^{(0)}(1|u), \quad   
T_{m}^{(1)}(u)=G_{m}^{(1)}(1|u), \nonumber \\ 
T_{m}^{(a)}(u)&=&T_{m}^{a}(1|u) ,\quad  2 \le a \le r-2, \label{sol-D1} \\
T_{m}^{(r-1)}(u)&=&G_{m}^{(r-1)}(1|u), \quad   
T_{m}^{(r)}(u)=G_{m}^{(r)}(1|u) \nonumber 	
\end{eqnarray} 
solves the functional equations (\ref{t-sysD1-1}-\ref{t-sysD1-6}) 
under the relations (\ref{rel5}) and (\ref{rel6}) for $k=1$.
\end{theorem}
\begin{theorem}({\it The $A_{2r}^{(2)}$ case, $r \ge 2$})
For $m\in Z_{\ge 0}$, 
\begin{eqnarray}
T_{m}^{(0)}(u)&=&F_{m}^{(0)}(2|u), \nonumber \\  
T_{m}^{(a)}(u)&=&T_{m}^{a}(2|u) ,\quad 1 \le a \le r-1, \label{sol-A2} \\
T_{2m}^{(r)}(u)&=&T_{m}^{r}(2|u), \quad  
T_{2m+1}^{(r)}(u)=H_{m+1}^{(r)}(2|u) \nonumber 
\end{eqnarray} 
solves the functional equations (\ref{t-sysA2-1}-\ref{t-sysA2-9}) under
 the relations (\ref{rel2}) and (\ref{rel3}) for $k=2$.
\end{theorem}
\begin{theorem}({\it The $A_{2r-1}^{(2)}$case, $r \ge 3$})
For $m\in Z_{\ge 0}$, 
\begin{eqnarray}
T_{m}^{(0)}(u)&=&G_{m}^{(0)}(2|u), \quad   
T_{m}^{(1)}(u)=G_{m}^{(1)}(2|u) ,\nonumber \\ 
T_{m}^{(a)}(u)&=&T_{m}^{a}(2|u) ,\quad 2 \le a \le r-1, \label{sol-A22}\\
T_{m}^{(r)}(u)&=&F_{m}^{(r)}(2|u) \nonumber 	
\end{eqnarray} 
solves the functional equations (\ref{t-sysA22-1}-\ref{t-sysA22-8}) 
under the relations (\ref{rel4}) and (\ref{rel5}) for $k=2$.
\end{theorem}
\begin{theorem}({\it The $D_{r+1}^{(2)}$ case, $r \ge 2$})
For $m\in Z_{\ge 0}$, 
\begin{eqnarray}
T_{2m}^{(0)}(u)&=&T_{m}^{0}(1|u), \quad  
T_{2m+1}^{(0)}(u)=H_{m+1}^{(0)}(1|u),\nonumber \\  
T_{m}^{(a)}(u)&=&T_{m}^{a}(1|u) ,\quad 1 \le a \le r-1, \label{sol-D2}\\
T_{2m}^{(r)}(u)&=&T_{m}^{r}(1|u), \quad  
T_{2m+1}^{(r)}(u)=H_{m+1}^{(r)}(1|u) \nonumber 	
\end{eqnarray} 
solves the functional equations (\ref{t-sysD2-1}-\ref{t-sysD2-8})
 under the relations (\ref{rel1}) and (\ref{rel2}) for $k=1$.
\end{theorem}
Remark1 :  
In eqs.(\ref{sol-A1})-(\ref{sol-D2}), the following type of boundary 
conditions are valid.
\[ {\rm For} \quad \alpha \le a \le \beta , \quad 
{\cal T}^{a}(u)=T_{1}^{(a)}(u) \]
where
$(\alpha,\beta)=(0,r)$ for (\ref{sol-A1});
$(\alpha,\beta)=(2,r-1)$ for (\ref{sol-B1}); 
$(\alpha,\beta)=(0,r)$ for (\ref{sol-C1});
$(\alpha,\beta)=(2,r-2)$ for (\ref{sol-D1}); 
$(\alpha,\beta)=(0,r-1)$ for (\ref{sol-A2}); 
$(\alpha,\beta)=(2,r)$ for (\ref{sol-A22}); 
$(\alpha,\beta)=(1,r-1)$ for (\ref{sol-D2}).  \\ 
Remark2 : 
In eq.(\ref{sol-A1}), if $ T_{1}^{(a)}(u)$ does not depend on 
$u \in {\bf C}$ for all $a$, then the 
following truncation holds: 
\[ T_{g}^{(b)}(u)=0 \qquad {\rm for} \quad {\rm any} \quad b. \] 
Now we enumerate lemmas that are necessary for the proofs of the theorems. 
\begin{lemma} \label{lemma1}
For $m \in {\bf Z }_{\ge 0}$ and $u \in {\bf C}$,
 $F_m^{(0)}(k|u)$ (\ref{F0})  (resp., $F_m^{(r)}(k|u)$ (\ref{Fr}))
 satisfy the following relations for $\alpha =0$ (resp., $\alpha =r$) 
 under the relation (\ref{rel3}) (resp., (\ref{rel4})).
\begin{eqnarray}
F_{m}^{(\alpha)}(k|u-1/k)F_{m}^{(\alpha)}(k|u+1/k)&=&T_{2m}^{\alpha}(k|u) ,\\
F_{m}^{(\alpha)}(k|u)F_{m+1}^{(\alpha)}(k|u)&=&T_{2m+1}^{\alpha}(k|u).
\end{eqnarray}
\end{lemma}
\begin{lemma} \label{lemma2}
For $m \in {\bf Z }_{\ge 1}$, $G_m^{(1)}(k|u)$ (\ref{G1}) and 
 $G_m^{(0)}(k|u)$ (\ref{G0}) (resp., $G_m^{(r-1)}(k|u)$ (\ref{Gr-1}) and 
 $G_m^{(r)}(k|u)$ (\ref{Gr}) ) satisfy the following relations for 
$(p,\alpha ,\beta)=(1,1,0)$ (resp., $(p,\alpha ,\beta)=(2,r-1,r)$) 
under the relation (\ref{rel5}) (resp., (\ref{rel6})). 
\begin{eqnarray}
\hspace{-45pt} &&
 G_{m}^{(\alpha)}(k|u+1/k) G_{m-1}^{(\beta)}(k|u) \nonumber \\
\hspace{-45pt} &&    = \left\{
          \begin{array}{ll}
            \det[{\cal S}_{m+1}^{[p]}
            \left[
              \begin{array}{c}
                m+1 \\
                1
              \end{array}
            \right]
            (k|u+(-m+2)/k)] & 
            {\rm for} \quad  m \in 2{\bf Z }_{\ge 1} \\
             \det[{\cal S}_{m+2}^{[p]}
            \left[
              \begin{array}{cc}
               2 & m+2 \\
               1 & 2
              \end{array}
            \right]
            (k|u-m/k)] & {\rm for} \quad m \in 2{\bf Z }_{\ge 0} +1,
      \end{array}
    \right.
 \label{relation3}
\end{eqnarray}
\begin{eqnarray}
\hspace{-45pt} && G_{m}^{(\alpha)}(k|u) G_{m}^{(\beta)}(k|u)=
    (-1)^m \det[{\cal S}_{m+1}^{[p]} 
           \left[
            \begin{array}{c}
             1 \\
             1
            \end{array}
           \right]
           (k|u+(-m+1)/k)],
 \label{relation1}
\end{eqnarray}
\begin{eqnarray}
\hspace{-45pt} && G_{m}^{(\alpha)}(k|u+1/k) G_{m+1}^{(\beta)}(k|u)=
    (-1)^{m+1} \det[{\cal S}_{m+2}^{[p]}
           \left[
            \begin{array}{c}
             1 \\
             2
            \end{array}
           \right]
           (k|u-m/k)],
 \label{relation4}
\end{eqnarray}
\begin{eqnarray}
\hspace{-45pt} && G_{m-1}^{(\alpha)}(k|u) G_{m}^{(\alpha)}(k|u+1/k)
\\ \nonumber 
\hspace{-45pt} && \hspace{70pt}=
 (-1)^{m} \det[{\cal S}_{m+2}^{[p]}
           \left[
            \begin{array}{cc}
             1 & 2 \\
             2 & m+2
            \end{array}
           \right]
           (k|u-m/k)],
 \label{relation6}
\end{eqnarray}
\begin{eqnarray}
\hspace{-45pt} && G_{m-1}^{(\beta)}(k|u-1/k) G_{m}^{(\beta)}(k|u)
\nonumber \\
\hspace{-45pt} && \hspace{70pt}=
    (-1)^{m} \det[{\cal S}_{m+1}^{[p]}
           \left[
            \begin{array}{c}
             1 \\
             m+1
            \end{array}
           \right]
           (k|u+(-m+1)/k)],
 \label{relation7}
\end{eqnarray}
\begin{eqnarray}
\hspace{-45pt} && G_{m}^{(\alpha)}(k|u) G_{m-1}^{(\beta)}(k|u+1/k) 
\\ \nonumber 
\hspace{-45pt} &&\hspace{70pt} 
  =  (-1)^{m} \det[{\cal S}_{m+2}^{[p]}
           \left[
            \begin{array}{cc}
             1 & 2 \\
             2 & 3
         \end{array}
           \right]
           (k|u+(-m-1)/k)].
 \label{relation8}
\end{eqnarray}
\end{lemma}
Remark3 : Lemma \ref{lemma1} for $\alpha =r, k=1$ with different boundary 
condition for ${\cal T}^{a}(u)$ was firstly proved in ref.\cite{KNH}. \\ 
Remark4 : Lemma \ref{lemma2} for $(p,\alpha ,\beta,k)=(2,r-1,r,1)$ with 
different boundary condition(\ref{boun}) for ${\cal T}^{a}(u)$ 
was firstly proved in ref.\cite{TK}. \\ 
The proofs of the theorems and lemmas in this section are quite similar to 
those in earlier work on solutions of the T-system
 \cite{KNH,KOS,TK}.
 That is, most of them reduce to the Jacobi 
identity:
\begin{eqnarray*}
\det {\cal M}\left[
   \begin{array}{c}
        b \\
        b 
   \end{array}
  \right]
 \det {\cal M}
   \left[
   \begin{array}{c}
        c \\ 
        c 
   \end{array}
  \right]-
 \det {\cal M}\left[
   \begin{array}{c}
        b \\
        c 
   \end{array}
  \right]
 \det {\cal M}\left[
   \begin{array}{c}
        c \\
        b 
   \end{array}
  \right]&=&
 \det {\cal M}\left[
   \begin{array}{cc}
        b & c\\
        b & c
   \end{array}
  \right]
   \det {\cal M}, \\ 
&&    (b \neq c) .   
\end{eqnarray*}
 So we shall not go into the detailed proofs here.
\eqreset
\section{Summary and Discussion}
In this paper, we have given the generalization of a 
 discretized Toda equation. Although the origin of it goes 
 back to the T-system, 
 a family of transfer matrix functional equations,  
 it has nothing to do with the solvable lattice models 
 to the author\symbol{39}  knowledge.   
It can be looked upon as a discretized affine Toda field 
equation. Solving it recursively, we have given, 
for $A_{n}^{(1)}$, $B_{n}^{(1)}$, $C_{n}^{(1)}$, $D_{n}^{(1)}$, 
$A_{n}^{(2)}$ and $D_{n+1}^{(2)}$, its solutions in terms of 
determinants or pfaffians.  
 
From the point of view of representation theory, the solutions of the 
 T-system \cite{KOS,KNH,TK} are considered to be Yangian analogue 
of the Jacobi-Trudi formulae, that is, Yangian character formulae in 
terms of fundamental representations. Then, what kind of underlying 
structures our new solutions here indicating ? 
Whether we can understand them within the framework of Yangian (or 
quantum affine algebras) or not is an open problem. 
There are related papers refs. \cite{KOS,TK,T,KLWZ,Wi,Z1,Z2}.

There remain problems to solve the functional equations associated with 
other Kac-Moody Lie algebras.
\\ 
{\bf Acknowledgments} \\ 
The author would like to 
thank Prof. A. Kuniba for useful discussions and comments on 
the manuscript, Prof. R. Hirota and Dr. S. Tsujimoto for kind interests 
in this work.
\eqreset
\setcounter{section}{0}
\def\thesection{\Alph{section}}
\renewcommand{\theequation}{\Alph{section}.\arabic{equation}}
\section{List of functional equations for other affine Lie algebras}
\def\thefootnote{}
 \footnote{We had removed this appendix from the 
 published version.}
 \def\thefootnote{\arabic{footnote}}
 \setcounter{footnote}{0}
For $G_{2}^{(1)} $, 
\begin{eqnarray}
\hspace{-45pt} &&     T_{m}^{(0)}(u-1) T_{m}^{(0)}(u+1)  =  
        T_{m+1}^{(0)}(u) T_{m-1}^{(0)}(u)+
        T_{m}^{(1)}(u)   \\ 
\hspace{-45pt} &&     T_{m}^{(1)}(u-1) T_{m}^{(1)}(u+1)  =  T_{m+1}^{(1)}(u) 
        T_{m-1}^{(1)}(u)+
        T_{m}^{(0)}(u) T_{3m}^{(2)}(u) \\
\hspace{-45pt} &&      T_{3m}^{(2)}(u-1/3) T_{3m}^{(2)}(u+1/3)   \nonumber \\
 \hspace{-45pt} && \hspace{50pt} =  T_{3m+1}^{(2)}(u) T_{3m-1}^{(2)}(u)  
       + T_{m}^{(1)}(u-2/3) T_{m}^{(1)}(u) T_{m}^{(1)}(u+2/3) \\ 
\hspace{-45pt} &&      T_{3m+1}^{(2)}(u-1/3) T_{3m+1}^{(2)}(u+1/3) 
\nonumber \\  
\hspace{-45pt} && \hspace{50pt}=T_{3m+2}^{(2)}(u) T_{3m}^{(2)}(u)+
        T_{m}^{(1)}(u-1/3) T_{m}^{(1)}(u+1/3) T_{m+1}^{(1)}(u) \\ 
\hspace{-45pt} &&      T_{3m+2}^{(2)}(u-1/3) T_{3m+2}^{(2)}(u+1/3)
\nonumber \\  
\hspace{-45pt} && \hspace{40pt}=T_{3m+3}^{(2)}(u) T_{3m+1}^{(2)}(u)+
        T_{m}^{(1)}(u) T_{m+1}^{(1)}(u-1/3) T_{m+1}^{(1)}(u+1/3) 
\end{eqnarray} 
\quad \\ 
 For $F_{4}^{(1)} $,
 \begin{eqnarray}
\hspace{-45pt} &&     T_{m}^{(0)}(u-1) T_{m}^{(0)}(u+1)  =  
        T_{m+1}^{(0)}(u) T_{m-1}^{(0)}(u)+
        T_{m}^{(1)}(u)   \\ 
\hspace{-45pt} &&     T_{m}^{(1)}(u-1) T_{m}^{(1)}(u+1)  =  T_{m+1}^{(1)}(u) 
        T_{m-1}^{(1)}(u)+
        T_{m}^{(0)}(u) T_{m}^{(2)}(u) \\
\hspace{-45pt} &&   T_{m}^{(2)}(u-1) T_{m}^{(2)}(u+1)  =  T_{m+1}^{(2)}(u) 
        T_{m-1}^{(2)}(u)+
        T_{m}^{(1)}(u) T_{2m}^{(3)}(u) \\
\hspace{-45pt} &&      T_{2m}^{(3)}(u-1/2) T_{2m}^{(3)}(u+1/2)  \nonumber \\  
\hspace{-45pt} && \hspace{50pt}
 = T_{2m+1}^{(3)}(u) T_{2m-1}^{(3)}(u)+
        T_{m}^{(2)}(u-1/2) T_{m}^{(2)}(u+1/2) T_{2m}^{(4)}(u) \\ 
\hspace{-45pt} &&      T_{2m+1}^{(3)}(u-1/2) T_{2m+1}^{(3)}(u+1/2) 
\nonumber \\  
\hspace{-45pt} &&\hspace{50pt}  
= T_{2m+2}^{(3)}(u) T_{2m}^{(3)}(u)+
        T_{m}^{(2)}(u) T_{m+1}^{(2)}(u) T_{2m+1}^{(4)}(u) \\ 
\hspace{-45pt} &&      T_{m}^{(4)}(u-1/2) T_{m}^{(4)}(u+1/2)   
       = T_{m+1}^{(4)}(u) T_{m-1}^{(4)}(u)+T_{m}^{(3)}(u) 
\end{eqnarray} 
\quad \\ 
 For $E_{6}^{(1)} $,
 \begin{eqnarray}
\hspace{-45pt} &&     T_{m}^{(0)}(u-1) T_{m}^{(0)}(u+1)  =  T_{m+1}^{(0)}(u) 
     T_{m-1}^{(0)}(u)+ T_{m}^{(6)}(u) \\
\hspace{-45pt} &&  T_{m}^{(a)}(u-1) T_{m}^{(a)}(u+1)  =  
     T_{m+1}^{(a)}(u) T_{m-1}^{(a)}(u)+
     T_{m}^{(a-1)}(u) T_{m}^{(a+1)}(u),\\
 \hspace{-45pt} &&\hspace{145pt} a=2,4  \nonumber \\ 
\hspace{-45pt} &&     T_{m}^{(1)}(u-1) T_{m}^{(1)}(u+1)  =  T_{m+1}^{(1)}(u) 
     T_{m-1}^{(1)}(u)+ T_{m}^{(2)}(u) \\
\hspace{-45pt} &&     T_{m}^{(3)}(u-1) T_{m}^{(3)}(u+1)  =  
     T_{m+1}^{(3)}(u) T_{m-1}^{(3)}(u)+
     T_{m}^{(2)}(u) T_{m}^{(4)}(u) T_{m}^{(6)}(u) \\
\hspace{-45pt} &&     T_{m}^{(5)}(u-1) T_{m}^{(5)}(u+1)  =  T_{m+1}^{(5)}(u) 
     T_{m-1}^{(5)}(u)+ T_{m}^{(4)}(u) \\
\hspace{-45pt} &&     T_{m}^{(6)}(u-1) T_{m}^{(6)}(u+1)  =  T_{m+1}^{(6)}(u) 
     T_{m-1}^{(6)}(u)+ T_{m}^{(0)}(u) T_{m}^{(3)}(u)  
\end{eqnarray} 
\quad \\ 
 For $E_{7}^{(1)} $,
 \begin{eqnarray}
\hspace{-45pt} &&     T_{m}^{(0)}(u-1) T_{m}^{(0)}(u+1)  =  T_{m+1}^{(0)}(u) 
     T_{m-1}^{(0)}(u)+ T_{m}^{(1)}(u) \\
\hspace{-45pt} &&  T_{m}^{(a)}(u-1) T_{m}^{(a)}(u+1)  =  
     T_{m+1}^{(a)}(u) T_{m-1}^{(a)}(u)+
     T_{m}^{(a-1)}(u) T_{m}^{(a+1)}(u),\\
      \hspace{-45pt} &&\hspace{145pt} a=1,2,4,5 \nonumber  \\ 
\hspace{-45pt} &&     T_{m}^{(3)}(u-1) T_{m}^{(3)}(u+1)  =  
     T_{m+1}^{(3)}(u) T_{m-1}^{(3)}(u)+
     T_{m}^{(2)}(u) T_{m}^{(4)}(u) T_{m}^{(7)}(u) \\
\hspace{-45pt} &&     T_{m}^{(6)}(u-1) T_{m}^{(6)}(u+1)  =  T_{m+1}^{(6)}(u) 
     T_{m-1}^{(6)}(u)+ T_{m}^{(5)}(u) \\
\hspace{-45pt} &&     T_{m}^{(7)}(u-1) T_{m}^{(7)}(u+1)  =  T_{m+1}^{(7)}(u) 
     T_{m-1}^{(7)}(u)+ T_{m}^{(3)}(u)   
\end{eqnarray} 
\quad \\ 
 For $E_{8}^{(1)} $,
 \begin{eqnarray}
\hspace{-45pt} &&     T_{m}^{(0)}(u-1) T_{m}^{(0)}(u+1)  =  T_{m+1}^{(0)}(u) 
     T_{m-1}^{(0)}(u)+ T_{m}^{(7)}(u) \\
\hspace{-45pt} &&  T_{m}^{(a)}(u-1) T_{m}^{(a)}(u+1)  =  
     T_{m+1}^{(a)}(u) T_{m-1}^{(a)}(u)+
     T_{m}^{(a-1)}(u) T_{m}^{(a+1)}(u),\\ 
      \hspace{-45pt} &&\hspace{145pt}  a=2,4,5,6  \nonumber \\ 
\hspace{-45pt} &&     T_{m}^{(1)}(u-1) T_{m}^{(1)}(u+1)  =  T_{m+1}^{(1)}(u) 
     T_{m-1}^{(1)}(u)+ T_{m}^{(2)}(u) \\
\hspace{-45pt} &&     T_{m}^{(3)}(u-1) T_{m}^{(3)}(u+1)  =  
     T_{m+1}^{(3)}(u) T_{m-1}^{(3)}(u)+
     T_{m}^{(2)}(u) T_{m}^{(4)}(u) T_{m}^{(8)}(u) \\
\hspace{-45pt} &&     T_{m}^{(7)}(u-1) T_{m}^{(7)}(u+1)  =  T_{m+1}^{(7)}(u) 
     T_{m-1}^{(7)}(u)+ T_{m}^{(0)}(u) T_{m}^{(6)}(u) \\
\hspace{-45pt} &&     T_{m}^{(8)}(u-1) T_{m}^{(8)}(u+1)  =  T_{m+1}^{(8)}(u) 
     T_{m-1}^{(8)}(u)+ T_{m}^{(3)}(u)   
\end{eqnarray} 
\quad \\ 
 For $E_{6}^{(2)} $,
 \begin{eqnarray}
\hspace{-45pt} &&     T_{m}^{(0)}(u-1/2) T_{m}^{(0)}(u+1/2)  =  
        T_{m+1}^{(0)}(u) T_{m-1}^{(0)}(u)+
        T_{m}^{(1)}(u)   \\ 
\hspace{-45pt} &&     T_{m}^{(1)}(u-1/2) T_{m}^{(1)}(u+1/2)  = 
T_{m+1}^{(1)}(u) 
        T_{m-1}^{(1)}(u)+
        T_{m}^{(0)}(u) T_{m}^{(2)}(u) \\
\hspace{-45pt} &&      T_{2m}^{(2)}(u-1/2) T_{2m}^{(2)}(u+1/2)  \nonumber \\  
\hspace{-45pt} && \hspace{45pt}
 = T_{2m+1}^{(2)}(u) T_{2m-1}^{(2)}(u)+
        T_{2m}^{(1)}(u)T_{m}^{(3)}(u-1/2)T_{m}^{(3)}(u+1/2)  \\ 
\hspace{-45pt} &&      T_{2m+1}^{(2)}(u-1/2) T_{2m+1}^{(2)}(u+1/2) 
\nonumber \\  
\hspace{-45pt} &&\hspace{45pt} 
= T_{2m+2}^{(2)}(u) T_{2m}^{(2)}(u)+
        T_{2m+1}^{(1)}(u)T_{m}^{(3)}(u) T_{m+1}^{(3)}(u)  \\ 
\hspace{-45pt} &&        T_{m}^{(3)}(u-1) T_{m}^{(3)}(u+1)  =  
        T_{m+1}^{(3)}(u)T_{m-1}^{(3)}(u)+
        T_{2m}^{(2)}(u) T_{m}^{(4)}(u) \\
\hspace{-45pt} &&      T_{m}^{(4)}(u-1) T_{m}^{(4)}(u+1)   
       = T_{m+1}^{(4)}(u) T_{m-1}^{(4)}(u)+T_{m}^{(3)}(u) 
\end{eqnarray} 
\quad \\ 
 For $D_{4}^{(3)} $,
 \begin{eqnarray}
\hspace{-45pt} &&     T_{m}^{(0)}(u-1/3) T_{m}^{(0)}(u+1/3)  =  
        T_{m+1}^{(0)}(u) T_{m-1}^{(0)}(u)+
        T_{m}^{(1)}(u)   \\ 
\hspace{-45pt} &&      T_{3m}^{(1)}(u-1/3) T_{3m}^{(1)}(u+1/3)  \nonumber \\  
\hspace{-45pt} &&  = T_{3m+1}^{(1)}(u) T_{3m-1}^{(1)}(u)+
        T_{3m}^{(0)}(u)T_{m}^{(2)}(u-2/3) 
        T_{m}^{(2)}(u) T_{m}^{(2)}(u+2/3) \\ 
\hspace{-45pt} &&      T_{3m+1}^{(1)}(u-1/3) T_{3m+1}^{(1)}(u+1/3) 
\nonumber \\  
\hspace{-45pt} && =T_{3m+2}^{(1)}(u) T_{3m}^{(1)}(u)+
        T_{3m+1}^{(0)}(u)T_{m}^{(2)}(u-1/3) 
        T_{m}^{(2)}(u+1/3) T_{m+1}^{(2)}(u) \\ 
\hspace{-45pt} &&      T_{3m+2}^{(1)}(u-1/3) T_{3m+2}^{(1)}(u+1/3)   
        =T_{3m+3}^{(1)}(u) T_{3m+1}^{(1)}(u) \nonumber \\ 
 \hspace{-45pt} && +
        T_{3m+2}^{(0)}(u)T_{m}^{(2)}(u) 
        T_{m+1}^{(2)}(u-1/3) T_{m+1}^{(2)}(u+1/3)   \\ 
\hspace{-45pt} &&    T_{m}^{(2)}(u-1) T_{m}^{(2)}(u+1)  =  T_{m+1}^{(2)}(u) 
        T_{m-1}^{(2)}(u)+T_{3m}^{(1)}(u) 
\end{eqnarray} 

\end{document}